# On Mobile Bluetooth Tags

Dmitry Namiot
Lomonosov Moscow State University
Moscow, Russia
dnamiot@gmail.com

Manfred Sneps-Sneppe
Ventspils University College
Ventspils, Latvia
manfreds.sneps@gmail.com

*Abstract*—This paper presents a new approach for hyper-local data sharing and delivery on the base of discoverable Bluetooth nodes. Our approach allows customers to associate user-defined data with network nodes and use a special mobile application (context-aware browser) for presenting this information to mobile users in proximity. Alternatively, mobile services can request and share local data in M2M applications rely on network proximity. Bluetooth nodes in cars are among the best candidates for the role of the bearing nodes.

## I. INTRODUCTION

Our paper is devoted to so-called context-aware computing. An original paper, introduced the term 'context-aware' [1] refers to context as location, identities of nearby people and objects, and changes to those objects. In other words, context is any information that can be used to characterize the situation of an entity. An entity here is a person, place, or object that is considered relevant to the interaction between a user and an application. This definition includes the user and applications themselves. This description makes it easier for an application developer to enumerate the context for a given application scenario [2].

There are plenty of papers devoted to context-aware (ubiquitous) computing [3] [4]. The reasons are obvious. From the end-users (customers) point of view, context-aware is a deep customization (localization) for data. It is precisely tuned output for mobile applications. Context processing lets us decide what kind of information should be presented for our users right now at this place exactly. Crawling of context dictates what kind of calculation should be done, what kind of requests to the external services to be raised, etc.

Actually, all the calculations for mobile users should be context-aware (for the ideal application, of course).

As it is mentioned above, the context is practically everything we can measure. But of course, for the practical applications we need some classification. On the first hand, we need some metrics, of course. In any case, our potential context should be measurable. So, the typical candidates are sound, light, accelerometer, etc. Among all this variety is quite obvious position - network infrastructure. It means that data available for mobile users could depend on the current state of the wireless infrastructure. And this state is a well known fingerprint [5]. For example, Wi-Fi fingerprint is a set of Wi-Fi access points and the signal strength with which they were heard. Some mobile application can show data to mobile users depends on the current fingerprint. Let us present just one example. It is enough to see the "visible" Wi-Fi access point in order to conclude which building and which floor we are in for the most of the buildings in MSU campus. Note, we do not need the location here. The wireless network fingerprint is enough for detecting the place. Also, we do not need the connectivity for the detected nodes. Their visibility is enough. The same is true for Bluetooth. It is so called network proximity [6]. The place here is not a location, but a proximity to some network nodes.

Network nodes could be mobile. E.g., Wi-Fi access point (Bluetooth node) used in fingerprint could be created right on a mobile phone, for example [7]. And Bluetooth nodes could be in cars too. For in car installation at least one important problem (power supply) is solved. Shortly, this approach was presented in [8]. And this paper discusses the latest development for this application.

The rest of the paper is organized as follows. In section II we describe related works. Section III introduces our concept of Bluetooth Data Points, and Section IV describes the prototype.

## II. RELATED WORKS

On the first hand, we can mention here our own SpotEx (Spot Expert) approach [9]. SpotEx links user-defined data and Wi-Fi nodes (Wi-Fi access points). For doing this, SpotEx introduces an external database with user-defined rules (productions or if-then operators) related to the Wi-Fi access points. Typical examples of conditions in our rules are:

```
IF Access_Point with SSID Café IS
visible AND RSSI (signal strength) is
within the given interval THEN

{activate some predefined content for
mobile users}
```

So, this service contains the following components:

- database (data store) with the productions (rules) associated with Wi-Fi networks;

- rule editor. It is a web application (including mobile web) which lets users add (edit) rule-set, associated with Wi-Fi networks;

- mobile application (context-aware browser). This browser detects Wi-Fi networks, checks the current conditions against the database of productions and executes rules (activates content).

Context-aware browse, based on the existing rules and discovered network environment, makes the above-mentioned user-defined content to be visible for mobile users. In other words, the visibility of mobile content depends on the network context (Wi-Fi network environment). Mobile users will see content which is relevant to their location. And the relevance here is user-defined. Content's owner describes the conditions for activations in the form of network proximity rules. Figure 1 illustrates this.

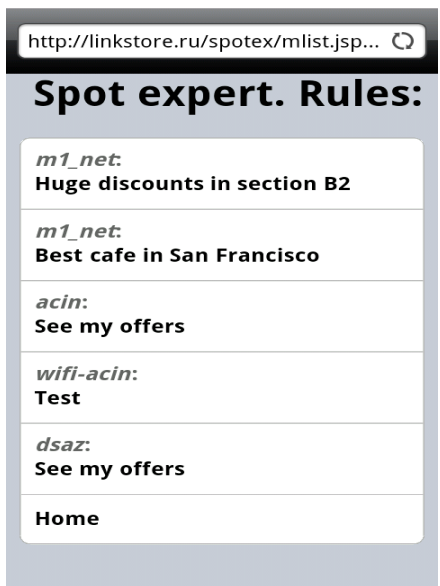

Fig. 1. SpotEx in-proximity rules

By the similar principles it could be done for Bluetooth nodes in the discoverable mode too.

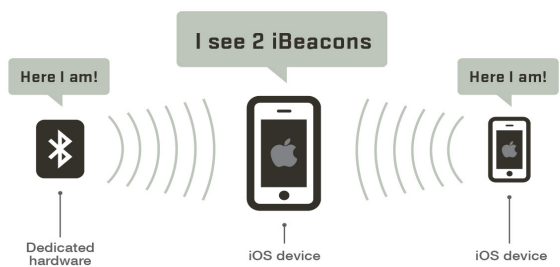

Fig. 2 iBeacons broadcasts [12]

The second example is an iBeacon from Apple [10]. The iBeacon is a tag with Bluetooth Low Energy chip [11]. The iBeacon simply broadcasts a presence message once per second to other devices within range of the Bluetooth radio (Figure 2).

It has a few identifying characteristics so that apps can distinguish the iBeacons they're interested in from a crowd [12]. Note that the iBeacon broadcasts have no data payload. iBeacons simply identify themselves via a UUID (unique identifier) and 2 numbers ("major" and "minor"). Originally (iOS platform) mobile application can only listen for specific UUIDs provided by the developer. On the Android platform mobile application can see a list of all iBeacons in proximity. The major and minor numbers could be configured to distinguish the tags within the particular application. Again, depends on the visible tags, mobile application can activate some content for mobile users (mobile devices).

Both approaches are very similar. For SpotEx the identification of the wireless node (e.g. MAC-address) plays the role of the UUID. And there is no need for the dedicated tag. In both cases, the content is separated from the network elements.

The paper [13] presents the concept of vehicular ad-hoc networks enables. The paper [14] discusses the effective data broadcasting. And our paper [15] discusses the role of Data Program Interfaces in mobile sensing.

### III. Bluetooth Data Points

Bluetooth Data Points (BDP) are Core Bluetooth nodes in discoverable mode which have some data associated with them. An idea for BDP was presented and described in the paper [16].

Let us see iBeacons core idea. We have some constant broadcast and a set of receivers. We need BLE just because broadcasted devices have no external power. But any Core Bluetooth node in discoverable mode is also a broadcaster. We can treat the MAC-address as a UUID. The power supply is not problem in case of cars. And Core Bluetooth makes this broadcast available for the widest set of devices.

We can even estimate the distance using RSSI (iBeacons use the similar approach – Figure 3). The standard approach is to measure the RSSI value at several prior-set distances between two phones in different environments. After that, a regression model could be constructed that depicts the relationship between RSSI level and distance. Moreover, we can estimate the uncertainty error when translating RSSI to distance.

For Bluetooth tag the distance estimation could be based on the ratio of the tag's signal strength (RSSI) over the calibrated transmitter power. The power is the known measured signal strength in RSSI at 1 meter away. Each tag (iBeacon in case of iOS) must be calibrated with this power value to allow the accurate distance estimation. The iBeacon output power is measured (calibrated) at a distance of 1 meter. Let's suppose that this is R1. The listening device will measure the RSSI of the device. Let's suppose it is R2. Since these numbers are in dBm, the ratio of the power is actually the difference in dB. So:

$$dBm\_ratio = R_1 - R_2 \qquad (1)$$

To convert that into a linear ratio, we use the standard formula:

$$linear\_ratio = 10^{(dBm\_ratio\,/\,10)} \qquad (2)$$

If we take into account the conservation of energy, then the signal strength must fall off as $1/r^2$ ($r$ here is a distance). So:

$$R = R_1 / r^2 \qquad (3)$$

$$r = \sqrt{linear\_ratio} \qquad (4)$$

Note, that, if our device is inside some building, then perhaps there will be internal reflections that make the signal decay slower than $1/r^2$. If the signal passes through a human body (water) then the signal will be attenuated. It's very likely that the antenna doesn't have equal gain in all directions. Metal objects in the room may create strange interference patterns.

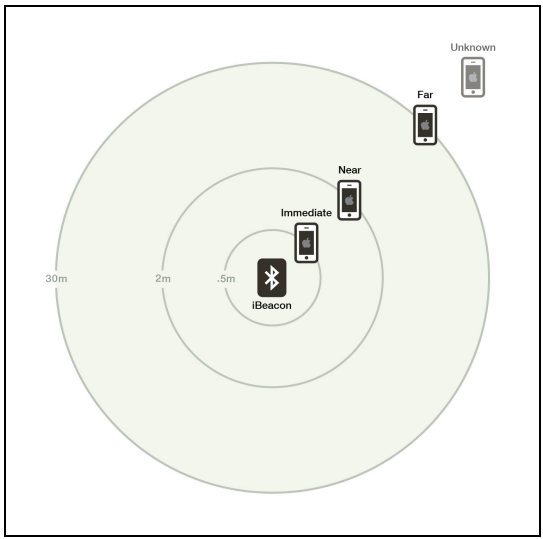

Fig. 3 RSSI based distance [12]

Now, we can associate some data with our broadcasters (Bluetooth nodes). It is similar to the location based systems. Their core is a database linked location (latitude and longitude pair) with any user-defined information. In BDP we can have a similar database, where MAC-addresses are linked with user-defined data. After that we have two main possibilities for mobile applications. Firstly, the application can obtain current fingerprint (get information about nearby nodes) and use it for obtaining local-related data from this database. Secondly, such a scan could be performed on the background and mobile user will get push-notifications with payloads related to local data [17]. Of course, mobile application with some user interface could be replaced with some automatic service. So, this approach will work for M2M applications too [18]. It is illustrated in Figure 4.

Let us provide yet another analogue for the main idea. BDP does not introduce a new network for cars, like presented in [19]. All data in this model are located and transported outside of Bluetooth nodes in cars. Bluetooth nodes play a role of presence sensors only.

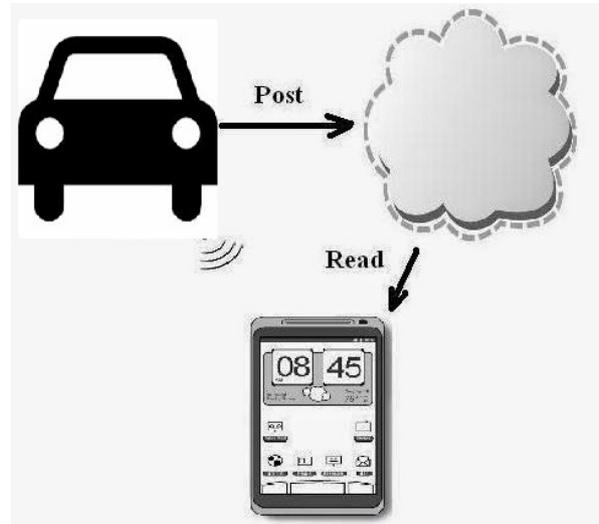

Fig. 4 Data Flow

The closest analogues are so called second screen applications [20]. It is about social networks for users of TV shows. Originally, it was planned to program them right on Smart TV. But nowadays, TV is just an initiator. The network for viewers exists in parallel on mobile devices carried by users anyway.

Conceptually, the data base for content has got a very simple structure. Data will be described individually for the each Bluetooth point. So, we have a key (MAC-address) and a vector of data chunks (texts, images, etc.) It is a typical key-value data model. This data model is one of the most used models for NoSQL approach. One of available examples of this project is Apache Accumulo [21]. It is distributed key-value store. Actually, the whole database for BDP could be a distributed hash table. The table rows as key-value pairs to provide a fast way to look up by a key item as attribute given by the value of a column qualifier of a row. In order to support lookups by more than one attribute of an entity, additional indexes can be built.

Data is represented as key-value pairs, where the key is comprised of the following elements: *RowID, Column* (*Family, Qualifier, Visibility*) and *Timestamp*. All elements of the Key and the Value are represented as byte arrays except for Timestamp, which is a Long. Accumulo sorts keys by element and lexicographically in ascending order. Timestamps are sorted in descending order so that later versions of the same Key appear first in a sequential scan. Tables consist of a set of sorted key-value pairs [22].

In terms of data design, BDP store the following information:

(recordID, MAC_address, data_array)

Each record describes a one data chunk (information element) for the given (MAC_address) Bluetooth node. Of course, we could have more than one information element for the same node.

If we think about statistics and performance, then each record should have, at least, two new fields: date when this element was created and last modified date when the record was last time modified. Also, we can allow users to switch on/off existing announces. Finally, we can propose the following structure for our records:

(recordID, MAC_address, timestamp_created, timestamp_modified, status, data_array)

here *status* field presents the Boolean (or the integer 1/0) value. It describes the current status for this data chunk.

Data array contains JSON structures. They describe content for the given data chunk. We choose JSON in order to keep the flexibility (easy to add a new data type) and move processing to the client side. JSON array, returned by the system, contains a list of obtained data chunks:

[
{
"type":"some_type","data":"some_data"},
{"type": ...},...
]

Note that the same kind of data will be returned by the programming API too. The field *type* here describes one of the standard types supported by the system. On the current stage, this system supports the following types: *text, url, image, email, phone, fbprofile, twprofile*. The *text* here is a sequence of chars, *url* and *image* describe some web-resource, *fbprofile* and *twprofile* are also web resources, but have a special meaning also. E.g., *fbprofile* is URL for some profile for Facebook, *twprofile* is the same for Twitter. It lets different programming clients decide by their own how to display (how to render) data. E.g., one client may render this as an ordinary hyperlink, the second may show a picture from the Facebook profile (obtained via public Facebook API), etc.

The typical query requests data by MAC_address. So, it is a direct scan via the primary index and it will be fast. The benchmark shows that Apache Accumulo can support very high levels of sustained throughputs of 100 million transactions per second [23].

As per collected statistics, the system can accumulate "browsing" events. An event here is the fact states that the device with address $MAC_1$ requests a data chunk provided by the device $MAC_2$ at the time *t*.

The basic algorithm as it is described above is very transparent. Our context-aware "browser" obtains a list of the visible Bluetooth node. Than for the each node we can perform database scan (lookup) and get data associated with this node. This request simply returns nothing in the case of Bluetooth nodes without associated data. All collected data could be packed in JSON array and this array will be returned back to the "browser". And the browser will perform data rendering. Nodes in the array could be sorted by the obtained RSSI (signal strength). In the normal case, most of the nearby Bluetooth nodes will be "empty" (they will be out of BDP circle). So, we can decrease the number of database lookups with some cache.

A Bloom filter is a method for representing a set of *n* elements $A = \{a_1, a_2,...,a_n\}$ also called keys to support membership queries.

The idea is to allocate a vector *V* of *m* bits, initially all set to 0, and then choose *k* independent hash functions, $K = \{h_1, h_2, ..., h_k\}$, each with a range $\{1,...,m\}$. For each element $a \in A$, the bits at positions $h_1(a), h_2(a), ..., h_k(a)$ in *V* are set to 1. Any particular bit might be set to 1 multiple times. Given a query for *b* (query is a key here) we check the bits at positions $h_1(b), h_2(b), ..., h_k(b)$. If any of them is 0, then certainly our key *b* is not in the set *A*. Otherwise, we conclude that b is in the set A. Keys here are MAC-addresses. The conclusion that some key is not in the cache lets us avoid extra request to database.

At the same time there is a certain probability for so called "false positive". In other words, we can make the false conclusion about presence MAC-address in the cache. The parameters *k* and *m* should be chosen such that the probability of a false positive (and hence a false hit to the database) is acceptable.

For Bloom filters we have a clear tradeoff between m and the probability of a false positive. Observe that after inserting n keys into a table of size m, the probability that a particular bit is still 0 is exactly

$$(1 - \frac{1}{m})^{kn} \qquad (5)$$

So, the probability of a false positive in this situation is

$$(1 - (1 - \frac{1}{m})^{kn})^k \qquad (6)$$

The latest expression could be approximated with this expression

$$(1 - e^{kn/m})^k \qquad (7)$$

It lets us, finally, to calculate false positive rate under various *m/n* and *k* combinations [24].

## IV. OUR PROTOTYPE

As a prototype, we've developed Android mobile application. It is available in Google Play [25]. It lets simulate the whole data flow. There are two parties (two mobile phones). One user can post some data to the cloud and link them to the own device. For posted data application starts (opens) Bluetooth node (right on poster's phone). So, posted data will be associated with the MAC-address of author's phone. It is illustrated in Figure 5.

The second party (parties) can start context-aware browser. This browser scans Bluetooth nodes in the proximity. For the each node it checks node's data on the cloud. It is the place for the above mentioned cache filter. In order to decrease the real hits to database, we check Bluetooth node against our cache. If the filter is positive, we can request (pull) data from our store.

By this way, browser accumulates all data chunks linked to the nearby Bluetooth nodes and shows them for user (Figure 6).

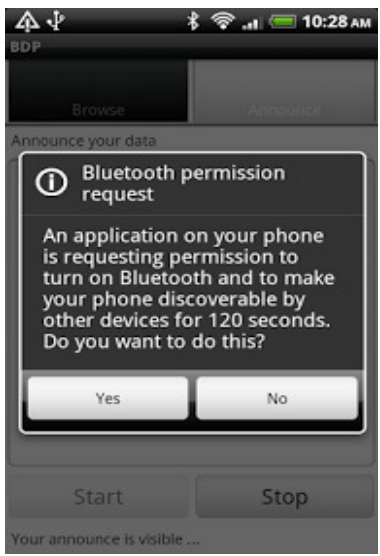

Fig. 5. Create BDP

As data chunks (linked data) we can use anything, that could be presented in HTML. So, it could be some URL, phone number (the browser makes it clickable), email, Twitter name, etc.

De-facto, browser uses REST based API for access to the database. This API returns obtained data as JSON array. Its interpretation is performed right in the applications. Of course, the same API could be used from another application. It is how M2M applications can work with BDP.

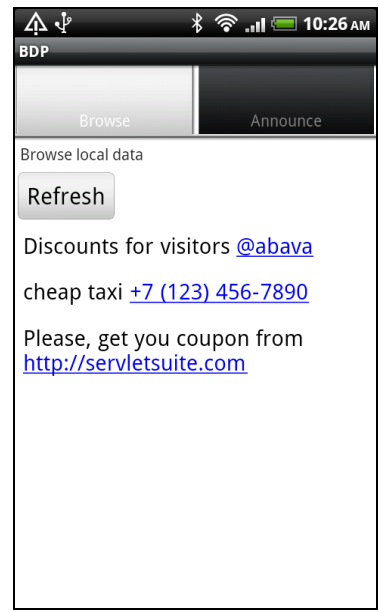

Fig. 6. BDP Browser

It is an open question (and subject to a separate research) - how to use the standard (modified) mobile browser as "context-aware" browser. Our initial experiments show, for example, that we can use a custom *WebView* control in Android OS for passing wireless context information right to JavaScript code. This feature opens the way for the incorporation of Bluetooth proximity processing right into web pages.

As per other context-aware models we can mention at least two use cases. On the first hand, it is localized communications (e.g., peer to peer chat and discussion groups). E.g., some mobile application can provide a link to the discussion group (forum) for all mobile users who can see a particular wireless network node. The link itself depends on the network node (e.g., its MAC-address). So, the discussion (chat) will be visible for those mobile users who are in the proximity to that node right now. It lets us "localize" communication channels. Users will chat (discuss) with pals in the physical proximity.

The next useful use case is associated with social streams. It is like a customized check-in [26]. Mobile users can link (associate) own profile on the social network (Facebook, Linkedin, etc.) with Bluetooth node on the own phone. So, as soon as Bluetooth node is on, mobile users with BDP "browser" explained above will be able to see the profiles for nearby users from the same social network. Think, for example about the conference, where visitors can advertise own profiles in Linkedin via Bluetooth.

And the classical example is, of course, some classified system. The author can link own announce to the mobile network node on the own phone. So, it is very easy to make it readable for the nearby mobile users. It could be done by switching on/off the mobile node (e.g., switch on/off a Bluetooth node, open/close Wi-Fi access point).

And user-defined content (classified) is traveling with the mobile phone (actually, with the author/owner). The content is available (readable) right there (and only there), where there is a mobile phone and where, consequently, is the author of the ads. And the author (owner) controls the visibility of own ad.

V. CONCLUSION

In this paper, we present a yet another approach for hyper-local data sharing and delivery on the base of discoverable Bluetooth nodes. Our discoverable data-hubs (Bluetooth Data Points) allow customers to associate any user-defined data with the programmatically created wireless network nodes. This process creates a distributed store of localized data. A special mobile application (context-aware browser) or service (in case of M2M applications) lets present this local information to mobile users or other services in proximity. For the mobile network nodes, associated data will "follow" them too. This data "movement" makes them available only in the proximity to the current location of data owner (creator).

ACKNOWLEDGMENT

We would like to thank prof. V.Vishnevsky for the valuable discussions.